\documentclass[prb,aps,twocolumn,showpacs,nobibnotes,epsf]{revtex4}

\usepackage{graphicx}
\usepackage{dcolumn}
\usepackage{bm}
\usepackage{SIunits}

\begin{document}
\title{Metamagnetic transition in Ca$_{1-x}$Sr$_x$Co$_2$As$_2$($x$ = 0 and 0.1) single crystals}
\author{J. J. Ying, Y. J. Yan, A. F. Wang, Z. J. Xiang, P. Cheng, G. J. Ye and X. H. Chen}
\altaffiliation{Corresponding
author} \email{chenxh@ustc.edu.cn} \affiliation{Hefei National
Laboratory for Physical Science at Microscale and Department of
Physics, University of Science and Technology of China, Hefei, Anhui
230026, People's Republic of China\\}

\begin{abstract}
We report the magnetism and transport measurements of CaCo$_2$As$_2$
and Ca$_{0.9}$Sr$_{0.1}$Co$_2$As$_2$ single crystals.
Antiferromagnetic transition was observed at about 70 K and 90 K for
CaCo$_2$As$_2$ and Ca$_{0.9}$Sr$_{0.1}$Co$_2$As$_2$, respectively.
Magnetism and magnetoresistance measurements reveal metamagnetic
transition from an antiferromagnetic state to a ferromagnetic state
with the critical field of 3.5 T and 1.5 T respectively along c-axis
for these two materials at low temperature. For the field along
ab-plane, spins can also be fully polarized above the field of 4.5 T
for Ca$_{0.9}$Sr$_{0.1}$Co$_2$As$_2$. While for CaCo$_2$As$_2$,
spins can not be fully polarized up to 7 T. We proposed the cobalt
moments of these two materials should be ordered ferromagnetically
within the ab-plane but antiferromagnetically along the
c-axis(A-type AFM).
\end{abstract}

\pacs{74.25.-q, 74.25.Ha, 75.30.-m}

\vskip 300 pt

\maketitle
\section{INTRODUCTION}
The layered ThCr$_2$Si$_2$ structure type is commonly observed for
$AT_2X_2$ compounds, in which $A$ is typically a  rare earth,
alkaline earth, or alkali element; $T$ is a transition-metal and $X$
is metalloid element. In this ThCr$_2$Si$_2$ structure, the $T_2X_2$
layers are made from edge-sharing $TX_4$ tetrahedra and $A$ ions are
intercalated between $T_2X_2$ layers. Novel physical properties were
observed in such ThCr$_2$Si$_2$ structure compounds including
magnetic ordering and superconductivity. Recent discoveries of high
temperature superconductivity in ThCr$_2$Si$_2$ structure pnictides
and selenides have led to renewed interest in this large class of
compounds\cite{Rotter, xlchen}. $A$Fe$_2$As$_2$ ($A$=Ca, Sr, Ba, Eu)
with the ThCr$_2$Si$_2$-type structure were widely investigated
because it is easy to grow large-size, high-quality single
crystals\cite{wugang, wangxf}. Superconductivity can be achieved
through doping or under high pressure. The maximum $T_{\rm c}$ for
the hole-doped samples is about 38 K and for the Co doped samples
the maximum $T_{\rm c}$ can reach 26 K\cite{Rotter, Sefat}.
Superconductivity up to 30 K was observed in $A_x$Fe$_{2-y}$Se$_2$
($A$=K, Rb, Cs and Tl) which also has the ThCr$_2$Si$_2$
structure\cite{xlchen, Mizuguchi, Wang, Ying, Krzton, Fang}. It is
very interesting to look for other materials with related structures
and investigate their physical properties to see if there can be
potential parent compounds for new high temperature superconductors.

$A$Co$_2$As$_2$ ($A$ is rare earth element) has the ThCr$_2$Si$_2$
structure, while their physical properties haven't been
systematically studied. The magnetic moment of the Co ion in this
type of material would not vanish due to the odd number of 3d
electron, thus we would anticipate an appearance of the magnetic
ordering phase. The magnetic properties of SrCo$_2$As$_2$ show
Curie-Weiss-like behavior and BaCo$_2$As$_2$ was reported as a
highly renormalized paramagnet in proximity to ferromagnetic
character\cite{Leithe, Sefat1}. LaOCoAs with the same CoAs layers
shows itinerant ferromagnetism\cite{Yanagi}. While CaCo$_2$P$_2$ was
reported as having ferromagnetically ordered Co planes, which are
stacked antiferromagnetically\cite{Reehuis}. The magnetic phase
diagram of Sr$_{1-x}$Ca$_x$Co$_2$P$_2$ is very complicated and
closely related to the structural changes\cite{Jia}. The magnetism
behavior in CoAs layered compounds is very interesting and it needs
further investigation.

In this article, we investigated the magnetism and transport
properties of CaCo$_2$As$_2$ and Ca$_{0.9}$Sr$_{0.1}$Co$_2$As$_2$
single crystals. Antiferromagnetic (AFM) ordering was observed below
$T_{\rm N}$ $\approx$ 70 K in CaCo$_2$As$_2$. And for
Ca$_{0.9}$Sr$_{0.1}$Co$_2$As$_2$, $T_{\rm N}$ increases to about 90
K. We determined the cobalt moments should be ordered
ferromagnetically within the ab-plane but antiferromagnetically
along the c-axis(A-type AFM). Metamagnetic transition corresponds to
a spin-flop transition from an AF to a FM state was observed in
these two samples.

\section{EXPERIMENTAL DETAILS}
High quality single crystals with nominal composition
Ca$_{1-x}$Sr$_x$Co$_2$As$_2$(x=0 and 0.1) were grown by conventional
solid-state reaction using CoAs as self-flux\cite{Sefat1}. The CoAs
precursor was first synthesized from stoichiometric amounts of Co
and As inside the silica tube at 700 $\celsius$ for 24 h. Then, the
mixture with ratio of Ca:Sr:CoAs=1-x:x:4 was placed in an alumina
crucible, and sealed in an quartz tube. The mixture was heated to
1200 $\celsius$ in 6 hours and then kept at this temperature for 10
hours, and later slowly cooled down to 950 $\celsius$ at a rate of 3
$\celsius$/ hour. After that, the temperature was cooled down to
room temperature by shutting down the furnace. The shining platelike
Ca$_{1-x}$Sr$_x$Co$_2$As$_2$ crystals were mechanically cleaved. The
actual composition of the single crystals were characterized by the
Energy-dispersive X-ray spectroscopy (EDX). The actual doping levels
are almost the same as the nominal values. Resistivity was measured
using the Quantum Design PPMS-9 and Magnetic susceptibility was
measured using the Quantum Design SQUID-VSM.

\section{RESULTS AND DISCUSSION}

\subsection{Crystal structure and resistivity}
\begin{figure}[t]
\centering
\includegraphics[width=0.5\textwidth]{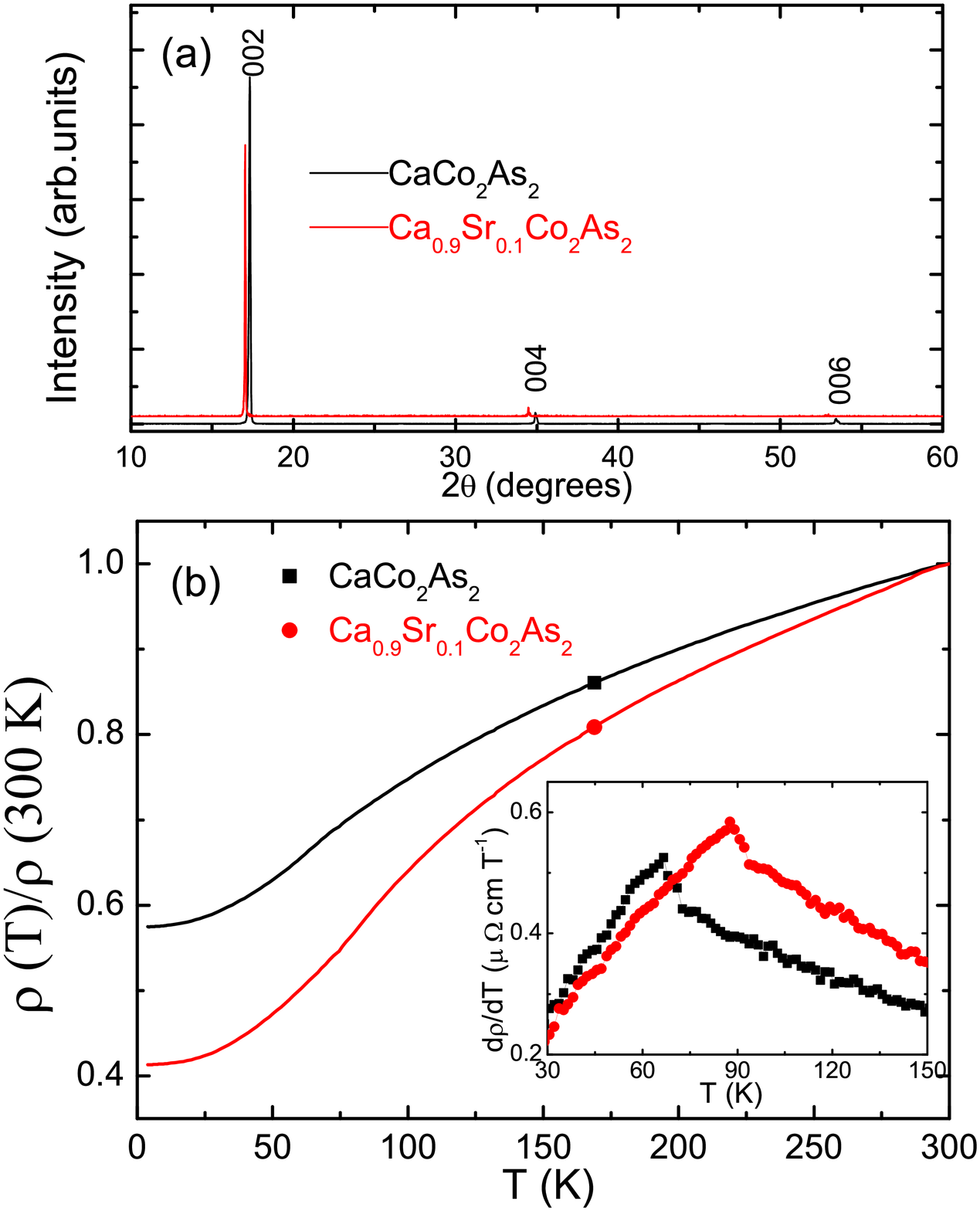}
\caption{(color online). (a): The single crystal x-ray diffraction
pattern of CaCo$_2$As$_2$ and Ca$_{0.9}$Sr$_{0.1}$Co$_2$As$_2$. (b):
Temperature dependence of in-plane resistivity for CaCo$_2$As$_2$
and Ca$_{0.9}$Sr$_{0.1}$Co$_2$As$_2$ single crystals. The inset was
the derivation of resistivity curves around $T_{\rm N}$.}
\label{fig1}
\end{figure}

Single crystals of Ca$_{1-x}$Sr$_x$Co$_2$As$_2$ ($x$ = 0 and 0.1)
were characterized by X-ray diffractions (XRD) using Cu $K_\alpha$
radiations as shown in Fig.1(a). Only (00$l$) diffraction peaks were
observed, suggesting uniform crystallographic orientation with the c
axis perpendicular to the plane of the single crystal. The c-axis
parameter was about 10.27 {\AA} for CaCo$_2$As$_2$ which is nearly
the same as the previous result\cite{Johnston}. For
Ca$_{0.9}$Sr$_{0.1}$Co$_2$As$_2$, c-axis parameter increases to
10.41 {\AA} due to the larger ion radius of Sr. The resistivity of
both samples shows metallic behavior as shown in Fig.1(b) which is
similar to SrCo$_2$As$_2$ and BaCo$_2$As$_2$\cite{Sefat, Leithe}.
The derivation of resistivity curves as shown in the inset of
Fig.1(b) show peaks at about 70 and 90 K for CaCo$_2$As$_2$ and
Ca$_{0.9}$Sr$_{0.1}$Co$_2$As$_2$, respectively. These temperatures
were consistent with the antiferromagnetic transition temperature
($T_{\rm N}$) which would be shown later in the magnetic
susceptibility measurements.

\subsection{Magnetic susceptibility and magnetoresistance}

\begin{figure}[t]
\centering
\includegraphics[width=0.5\textwidth]{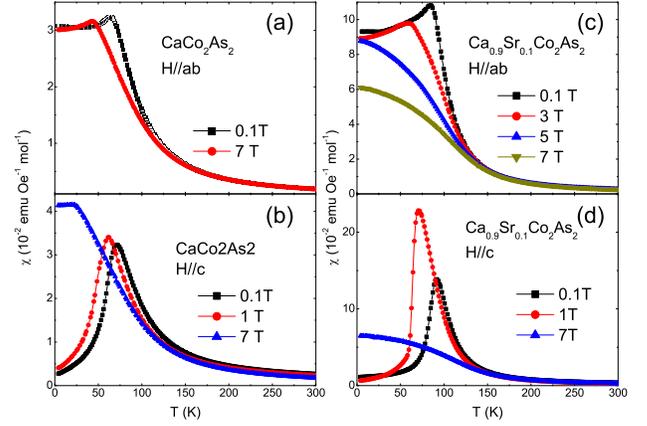}
\caption{(color online). Temperature dependence of the
susceptibility for CaCo$_2$As$_2$ with magnetic field along (a) and
perpendicular (b) to the ab-plane. Temperature dependence of the
susceptibility for Ca$_{0.9}$Sr$_{0.1}$Co$_2$As$_2$ with magnetic
field along (c) and perpendicular (d) to the ab-plane.} \label{fig2}
\end{figure}

\begin{figure}[t]
\centering
\includegraphics[width=0.5\textwidth]{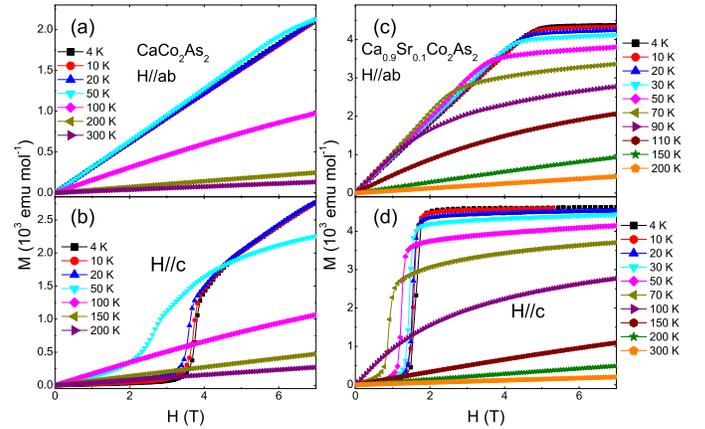}
\caption{(color online). Isothermal magnetization hysteresis of
CaCo$_2$As$_2$ with magnetic field along (a) and perpendicular (b)
to the ab-plane. MH curves for Ca$_{0.9}$Sr$_{0.1}$Co$_2$As$_2$ with
field along (c) and perpendicular (d) to the ab-plane at certain
temperature.} \label{fig3}
\end{figure}

\begin{figure}[t]
\centering
\includegraphics[width=0.5\textwidth]{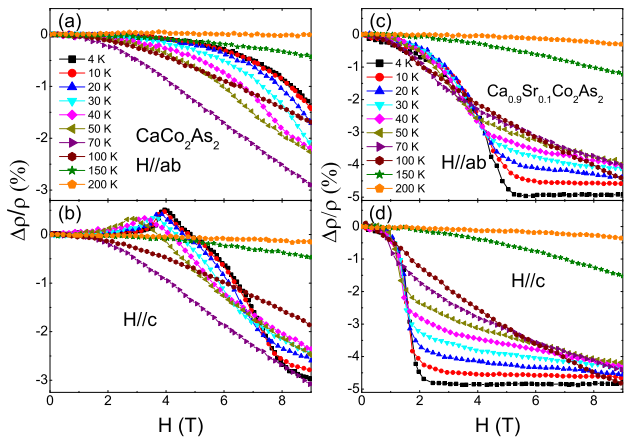}
\caption{(color online). Magnetoresistance of CaCo$_2$As$_2$ with
magnetic field up to 9T along (a) and perpendicular (b) to the
ab-plane. MR curves for Ca$_{0.9}$Sr$_{0.1}$Co$_2$As$_2$ with field
along (c) and perpendicular (d) to the ab-plane at certain
temperature.} \label{fig4}
\end{figure}

Figure 2 (a) and (b) show the temperature dependence of
susceptibility for CaCo$_2$As$_2$ under various magnetic field up to
7 T applied within ab-plane and along c-axis, respectively. The
susceptibility drops at around 70 K with the field of 0.1 T applied
along and perpendicular to the ab-plane, which indicates the
antiferromagnetic state below 70 K in this material. $T_{\rm N}$ was
suppressed by increasing the field and was suppressed to 50 K under
the field of 7 T applied along the ab-plane. However, with the field
of 7 T applied along the c-axis, the susceptibility saturated at low
temperature which indicated the ferromagnetic state at low
temperature. For Ca$_{0.9}$Sr$_{0.1}$Co$_2$As$_2$, $T_{\rm N}$
increases to 90 K. Under high magnetic field, the system changes to
ferromagnetic state with field applied both along and perpendicular
to the ab-plane as shown in Fig. 2 (c) and (d). These results
indicate that a metamagnetic transition from antiferromagnetism
(AFM) to ferromagnetism (FM) occurs with increasing magnetic field
at low temperature.

In order to further investigate such metamagnetic transition, we
performed the isothermal magnetization hysteresis measurement as
shown in Figure 3. The magnetization (M) almost increases linearly
for CaCo$_2$As$_2$ with the field H applied along the ab-plane at
various temperatures. The slope of MH curves is almost the same
below 50 K and decrease with increasing the temperature above
$T_{\rm N}$. While for the field applied along the c-axis, M
increases very sharply when H increases to about 3.5 T at 4 K. With
increasing the temperature, such behavior weakened and vanished
above $T_{\rm N}$. This behavior clearly indicates the spin
reorientation in CaCo$_2$As$_2$ at high magnetic field. While for
Ca$_{0.9}$Sr$_{0.1}$Co$_2$As$_2$, M increases very steeply with H
increasing to 1.5 T and saturates at high field with the field
applied along the c-axis at 4 K. The transition field gradually
decreases with increasing the temperature and vanishes above $T_{\rm
N}$. M increases almost linearly with H applied along ab-plane at
low field at 4 K which is similar to the CaCo$_2$As$_2$, further
increasing the field leads to the saturation of M. The values of the
saturated M are almost the same for the field along and
perpendicular to the ab-plane. This result clearly indicates that
almost all the magnetic ions can be tuned by the magnetic field
above a critical field. The critical magnetic field of
Ca$_{0.9}$Sr$_{0.1}$Co$_2$As$_2$ is lower than CaCo$_2$As$_2$, which
is probably due to the weakening of inter layer AFM coupling. Spins
can be tuned much easier with magnetic field applied along c-axis
than along ab-plane, for CaCo$_2$As$_2$, spins could not be aligned
along ab-plane even under 7 T.

We further measured the magnetoresistance (MR) of CaCo$_2$As$_2$ and
Ca$_{0.9}$Sr$_{0.1}$Co$_2$As$_2$ from 4 to 200 K up to 9 T as shown
in Figure 4. Magnetoresistance can be hardly detected at high
temperature and gradually became negative with decreasing the
temperature. For CaCo$_2$As$_2$, the magnitude of magnetoresistance
gradually increases above $T_{\rm N}$ and gradually decreases below
$T_{\rm N}$ with decreasing the temperature. This is because the
presence of the FM order tends to suppress the spin scattering.
Magnetic field gradually polarized the spins of Co ions with
decreasing the temperature above $T_{\rm N}$, while below the AFM
transition temperature, it became much more difficult to polarize
the spins with the temperature cooling down. The MR with H
perpendicular to the ab-plane was almost the same with H along
ab-plane above $T_{\rm N}$. While the temperature below $T_{\rm N}$,
MR became positive at low magnetic field. When the applied magnetic
field surpass the critical field of metamagnetic transition, MR
gradually becomes negative and its magnitude increases with
increasing the magnetic field. Similar MR behavior was observed
above $T_{\rm N}$ with field applied along and perpendicular to the
ab-plane. The MR changes greatly at the critical field of
metamagnetic transition. The metamagnetic transition of
Ca$_{0.9}$Sr$_{0.1}$Co$_2$As$_2$ is much sharper comparing with
CaCo$_2$As$_2$ with H applied along c-axis. MR became constant when
all the spins were tuned to FM state. The values of MR are almost
the same for the field along and perpendicular to the ab-plane which
indicates that all the spins can be tuned by high magnetic field.
Similar MR behaviors across the AF and the FM phase boundary were
also observed in other materials such as Na$_{0.85}$CoO$_2$, layered
ruthenates and colossal magnetoresistance materials\cite{Luo,
Nakatsuji}. The MR result about the metamagnetic transition is
consistent with the MH measurements in these two materials. The
magnetoresistance with field applied along and perpendicular to the
ab-plane are almost the same under high magnetic field which
indicates that magnetoresistance was mainly induced from magnetic
scattering from Co$^{2+}$ and all the Co$^{2+}$ spins can be tuned
above the critical field of metamagnetic transition at low
temperature.

\begin{figure}[t]
\centering
\includegraphics[width=0.5\textwidth]{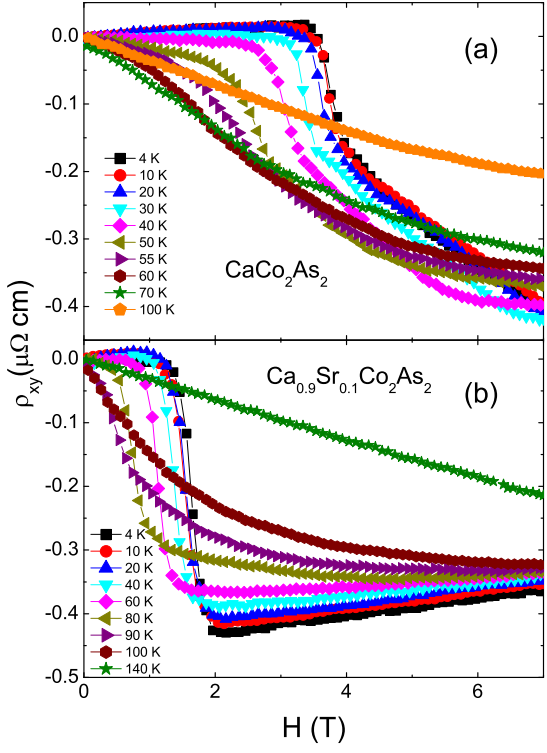}
\caption{(color online). Field dependence of Hall resistivity at
various temperatures for CaCo$_2$As$_2$ (a) and
Ca$_{0.9}$Sr$_{0.1}$Co$_2$As$_2$ (b).} \label{fig5}
\end{figure}

\begin{figure}[t]
\centering
\includegraphics[width=0.5\textwidth]{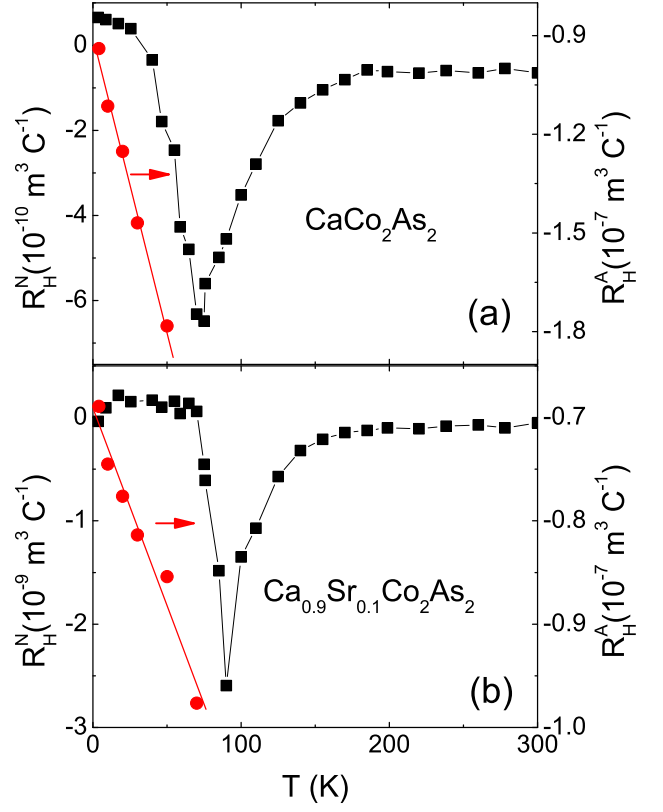}
\caption{(color online). The temperature dependence of anomalous
Hall coefficient $R^A_H$ and normal Hall coefficient $R^N_H$ for
CaCo$_2$As$_2$ (a) and Ca$_{0.9}$Sr$_{0.1}$Co$_2$As$_2$ (b).}
\label{fig6}
\end{figure}

\begin{figure}[t]
\centering
\includegraphics[width=0.5\textwidth]{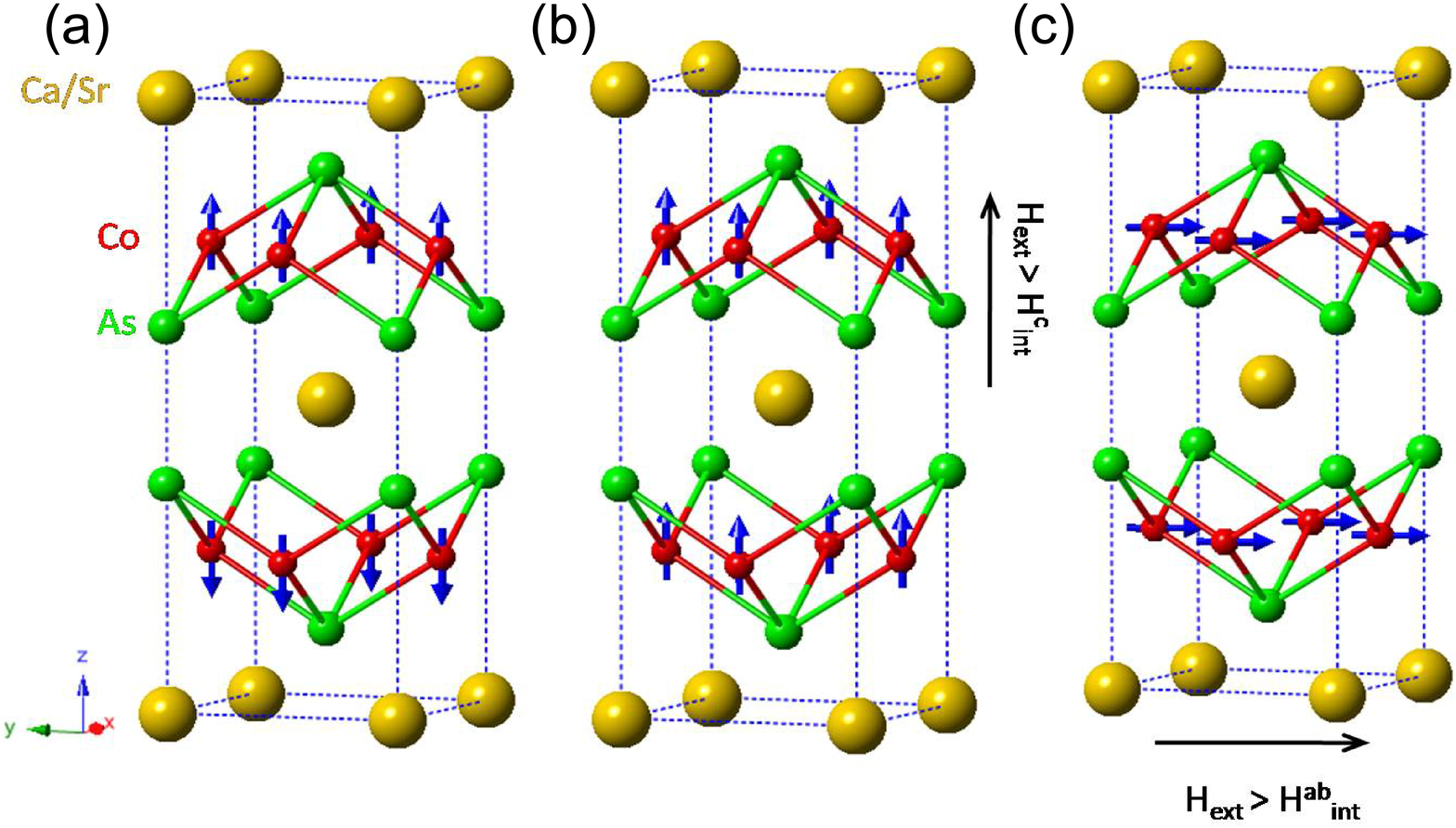}
\caption{(color online). (a): Possibly magnetic structure deduced
from magnetism and MR measurement. Magnetic structure above the
critical field with H along (b) and perpendicular (c) to c-axis.}
\label{fig7}
\end{figure}
\subsection{Hall coefficient and magnetic structure}

We also measured the Hall resistivity $\rho_{xy}$ of CaCo$_2$As$_2$
and Ca$_{0.9}$Sr$_{0.1}$Co$_2$As$_2$ as shown in Figure 5.
$\rho_{xy}$ is measured by sweeping field from -7 T to +7 T at
various temperature, thus the accurate Hall resistivity $\rho_H$ is
obtained by using $[\rho_{xy}(+H)-\rho_{xy}(-H)]/2$, where
$\rho_{xy}(\pm H)$ is $\rho_{xy}$ under positive or negative
magnetic field. We found $\rho_{xy}$ shows a steep decrease at
certain field $H_C$ below $T_{\rm N}$ similar to isothermal MR,
which arises from the jump magnitudes of magnetization \textbf{M}
due to the spin-flop of Co ions induced by external field
\textbf{H}. It is well known that Hall effect arises from two parts
of normal Hall effect and anomalous Hall effect in ferromagnetic
metals, in which anomalous Hall resistivity is proportional to the
magnetization \textbf{M}. Hall resistivity $\rho_{xy} = R^N_HH +
R^A_H4\pi M$, where $R^N_H$ is the normal Hall coefficient, and
$R^A_H$ is the anomalous Hall coefficient\cite{nagaosa}. We
extracted the $R^N_H$ from H-linear term of $\rho_{xy}$ at low
field. Temperature dependence of $R^N_H$ for CaCo$_2$As$_2$ and
Ca$_{0.9}$Sr$_{0.1}$Co$_2$As$_2$ are shown in Fig.6(a) and (b),
respectively. The value of $R^N_H$ is negative at high temperature,
indicating the primate carrier in these two materials is electron.
The magnitude of $R^N_H$ gradually increases with decreasing the
temperature and it reaches its maximum value at around $T_{\rm N}$.
Below $T_{\rm N}$, the magnitude of $R^N_H$ decreases very quickly
with decreasing the temperature. The anomalous Hall coefficient
$R^A_H$ below $T_{\rm N}$ is inferred from the ratio of the jump
magnitudes $\Delta M$ and $\Delta \rho_{xy}$ around metamagnetic
transition field. $R^A_H=\Delta\rho_{xy}/4\pi \Delta M$ is also
plotted in Fig.6 (a) and (b). The magnitude of $R^A_H$ decreases
linearly with decreasing temperature below $T_{\rm N}$. Such
behavior was also observed in TaFe$_{1+y}$Te$_3$ system\cite{Liu}.

Based on the above results of magnetic susceptibility, MR and Hall
resistivity measurement, possible magnetic structures for the spins
of Co ions are proposed as shown in Fig. 7 (a), (b) and (c). Below
$T_{\rm N}$, Co spins of these two materials should be ordered
ferromagnetically within the ab-plane but antiferromagnetically
along the c-axis(A-type AFM) under low magnetic field. When the
external field surpass the inner field \textbf{H$_{int}$} of Co
spins. All the Co spins aligned along the direction of H. The
external field H along ab-plane is much harder to tune the AFM
ordering of Co spins than with H along c-axis.

The AFM ordering found in CaCo$_2$As$_2$ and
Ca$_{0.9}$Sr$_{0.1}$Co$_2$As$_2$ is very different from their
isostructure compounds SrCo$_2$As$_2$ and BaCo$_2$As$_2$ in which no
magnetic ordering was observed. Such a difference might due to the
much smaller c/a ratio in CaCo$_2$As$_2$ than in SrCo$_2$As$_2$ and
BaCo$_2$As$_2$, similar properties were also observed in
CaCo$_2$P$_2$\cite{Jia}. High magnetic field can tune almost all the
spins, and metamagnetic transition was observed by increasing the
magnetic field at low temperature which was similar to
EuFe$_2$As$_2$\cite{wutao}. The magnetic property in this system is
strongly correlated to the structure. The critical field of
metamagnetic transition decreased by doping Sr, which was probably
due to the enlarged c-axis parameter and decreased the inter layer
AFM coupling.

\section{CONCLUSION}
In conclusion, we found AFM ordering in CaCo$_2$As$_2$ and
Ca$_{0.9}$Sr$_{0.1}$Co$_2$As$_2$ at 70 and 90 K, respectively. A
metamagnetic transition from AFM to FM occurs with increasing
magnetic field in these two compounds. Little Sr doping in
CaCo$_2$As$_2$ would effectively decrease the critical field of
metamagnetic transition. The magnetic susceptibility and MR
measurements all indicate that the cobalt moments of these two
materials should be ordered ferromagnetically within the ab-plane
but antiferromagnetically along the c-axis(A-type AFM), which is the
same with their isostructure compound CaCo$_2$P$_2$\cite{Reehuis,
Reehuis2}.

{\bf ACKNOWLEDGEMENT} This work is supported by the National Basic
Research Program of China (973 Program, Grant No. 2012CB922002 and
No. 2011CB00101), National Natural Science Foundation of China
(Grant No. 11190021 and No. 51021091), the Ministry of Science and
Technology of China, and Chinese Academy of Sciences.


\begin{references}
\bibitem{Rotter}
M. Rotter, M. Tegel, D. Johrendt, Phys. Rev. Lett. {\bf 101},
107006(2008).
\bibitem{xlchen}
J. Guo, S. Jin, G. Wang, S. Wang, K. Zhu, T. Zhou, M. He and X.
Chen, Phys. Rev. B {\bf 82}, 180520(R) (2010).
\bibitem{wugang}
G Wu, H Chen, TWu, Y L Xie, Y J Yan, R H Liu, X FWang, J J Ying and
X H Chen,  J. Phys.: Condens. Matter {\bf 20} 422201 (2008)
\bibitem{wangxf}
X. F. Wang, T. Wu, G. Wu, H. Chen, Y. L. Xie, J. J. Ying, Y. J. Yan,
R. H. Liu and X. H. Chen,  Phys. Rev. Lett. {\bf 102}, 117005(2009).
\bibitem{Sefat}
Athena S. Sefat, Rongying Jin, Michael A. McGuire, Brian C. Sales,
David J. Singh, and David Mandrus, Phys. Rev. Lett. {\bf 101},
117004 (2008).
\bibitem{Mizuguchi}
Yoshikazu Mizuguchi, Hiroyuki Takeya, Yasuna Kawasaki, Toshinori
Ozaki, Shunsuke Tsuda, Takahide Yamaguchi and Yoshihiko Takano,
Appl. Phys. Lett. 98, 042511 (2011).
\bibitem{Wang}
A. F. Wang, J. J. Ying, Y. J. Yan, R. H. Liu, X. G. Luo, Z. Y. Li,
X. F. Wang, M. Zhang, G. J. Ye, P. Cheng, Z. J. Xiang, X. H.
 Chen, Phys. Rev. B {\bf 83}, 060512(R) (2011).
\bibitem{Ying}
J. J. Ying, X. F. Wang, X. G. Luo, A. F. Wang, M. Zhang, Y. J. Yan,
Z. J. Xiang, R. H. Liu, P. Cheng, G. J. Ye, X. H. Chen , Phys. Rev.
B {\bf 83}, 212502 (2011).
\bibitem{Krzton}
A. Krzton-Maziopa, Z. Shermadini, E. Pomjakushina, V. Pomjakushin,
M. Bendele, A. Amato, R. Khasanov, H. Luetkens and K. Conder, J.
Phys.: Condens. Matter {\bf 23}, 052203 (2011).
\bibitem{Fang}
Minghu Fang, Hangdong Wang, Chiheng Dong, Zujuan Li, Chunmu Feng,
Jian Chen, H.Q. Yuan, EPL, {\bf 94}, 27009 (2011).
\bibitem{Sefat1}
A. S. Sefat, D. J. Singh, R. Jin, M. A. McGuire, B. C. Sales, and D.
Mandrus, Phys. Rev. B {\bf 79}, 024512 (2009).
\bibitem{Leithe}
A. Leithe-Jasper, W. Schnelle, C. Geibel, and H. Rosner, Phys. Rev.
Lett. {\bf 101}, 207004 (2008).
\bibitem{Yanagi}
H. Yanagi, R. Kawamura, T. Kamiya, Y. Kamihara, M. Hirano, T.
Nakamura, H. Osawa, and H. Hosono, Phys. Rev. B {\bf 77}, 224431
(2008).
\bibitem{Luo}
J. L. Luo, N. L.Wang, G.T. Liu, D. Wu, X. N. Jing, F. Hu, and T.
Xiang, Phys. Rev. Lett. {\bf 93}, 187203 (2004).
\bibitem{Nakatsuji}
S. Nakatsuji, D. Hall, L. Balicas, Z. Fisk, K. Sugahara, M.
Yoshioka, and Y. Maeno, Phys. Rev. Lett. {\bf 90}, 137202 (2003).
\bibitem{nagaosa}
N. Nagaosa, J. Sinova, S. Onoda, A. H. MacDonald, and N. P. Ong,
Rev. Mod. Phys. {\bf82}, 1539(2010).
\bibitem{Liu}
R. H. Liu, M. Zhang, P. Cheng, Y. J. Yan, Z. J. Xiang, J. J. Ying,
X. F. Wang, A. F. Wang, G. J. Ye, X. G. Luo, and X. H. Chen, Phys.
Rev. B {\bf 84}, 184432  (2011).
\bibitem{Jia}
Shuang Jia, A. J. Williams, P. W. Stephens, and R. J. Cava, Phys.
Rev. B {\bf 80}, 165107 (2009).
\bibitem{Johnston}
David C. Johnston, Advances in Physics {\bf 59}, 803¨C1061 (2010).
\bibitem{wutao}
T. Wu, G. Wu, H. Chen, Y. L. Xie, R. H. Liu, X. F. Wang and X. H.
Chen, J. Mag. Mag. Mat. {\bf 321}, 3870-3874 (2009).
\bibitem{Reehuis}
M. Reehuis and W. Jeitschko, J. Phys. Chem. Solids {\bf 51}, 961
(1990).
\bibitem{Reehuis2}
M. Reehuis, W. Jeitschko, G. Kotzyba, B. Zimmer, and X. Hu, J.
Alloys Compd. {\bf 266}, 54 (1998).
\newpage


\noindent

\end{references}
\end{document}